\documentclass[prl,amsmath,amssymb,twocolumn,superscriptaddress]{revtex4}

\usepackage{amsmath}
\usepackage{amssymb}
\usepackage{amsfonts}
\usepackage{color}
\usepackage{graphicx}

\newcommand{\RR}{\right}
\newcommand{\LL}{\left}
\newcommand{\m}{\mathrm}

\newcommand{\eref}[1]{Eq.~(\ref{#1})}
\newcommand{\fref}[1]{Fig.~\ref{#1}}

\begin{document}

\title{Single-photon cavity optomechanics mediated by a quantum two-level system}

\author{J.-M. Pirkkalainen}
\affiliation{Department of Applied Physics, Aalto University School of Science, P.O. Box 11100, FI-00076 Aalto, Finland}

\author{S.~U. Cho}
\thanks{Present address: Department of Physics and Astronomy, Seoul National University, Seoul 151-747, Korea}
\affiliation{O.~V.~Lounasmaa Laboratory, Low Temperature Laboratory, Aalto University, P.O. Box 15100, FI-00076 Aalto, Finland.}

\author{F. Massel}
\affiliation{Department of Physics, Nanoscience Center, University of Jyv\"askyl\"a, P.O. Box 35 (YFL)
FI-40014 University of Jyv\"askyl\"a, Finland}

\author{J. Tuorila}
\affiliation{Department of Physics, University of Oulu, FI-90014, Finland}

\author{T.~T. Heikkil\"a}
\affiliation{Department of Physics, Nanoscience Center, University of Jyv\"askyl\"a, P.O. Box 35 (YFL)
FI-40014 University of Jyv\"askyl\"a, Finland}

\author{P.~J.~Hakonen}
\affiliation{O.~V.~Lounasmaa Laboratory, Low Temperature Laboratory, Aalto University, P.O. Box 15100, FI-00076 Aalto, Finland.}

\author{M.~A. Sillanp\"a\"a}
\affiliation{Department of Applied Physics, Aalto University School of Science, P.O. Box 11100, FI-00076 Aalto, Finland}

\begin{abstract}
{\bf

Coupling electromagnetic waves in a cavity and mechanical vibrations via the radiation pressure of the photons \cite{KippenbergReview,OptoReview2014} is a promising platform for investigations of quantum mechanical properties of motion of macroscopic bodies and thereby the limits of quantum mechanics \cite{Teufel2011b,AspelmeyerCool11}. A drawback is that the effect of one photon tends to be tiny, and hence one of the pressing challenges is to substantially increase the interaction strength towards the scale of the cavity damping rate. A novel scenario is to introduce into the setup a quantum two-level system (qubit), which, besides strengthening the coupling, allows for rich physics via strongly enhanced nonlinearitites \cite{Kippenberg2013TLS,CiracTLS2013,LSETNEMSth,Treutlein2014}. Addressing these issues, here we present a design of cavity optomechanics in the microwave frequency regime involving a Josephson junction qubit. We demonstrate boosting of the radiation pressure interaction energy by six orders of magnitude, allowing to approach the strong coupling regime, where a single quantum of vibrations shifts the cavity frequency by more than its linewidth. We observe nonlinear phenomena at single-photon energies, such as an enhanced damping due to the two-level system. This work opens up nonlinear cavity optomechanics as a plausible tool for the study of quantum properties of motion. }
\end{abstract}

\maketitle

One of the most successful realizations of cavity optomechanics in the recent past has been that using microwave-regime superconducting cavities made with lithographic techniques \cite{LehnertNphOpto08,MechAmpPaper,Teufel2011b}. Here, the oscillatory displacement $x$ changes the effective capacitance $C(x)$ of the cavity and hence its resonance frequency $\omega_c$, and the radiation-pressure coupling energy is $g_0 \equiv  (\partial \omega_c /  \partial x )x_{\m{zp}}  = (\omega_c/2C)(\partial C/\partial x) x_{\m{zp}}$. Here,  $x_{\m{zp}}$ is the zero-point motion. The largest radiation pressure coupling values $g_0/2\pi \simeq 1$ MHz have been demonstrated in the optical frequency domain \cite{AspelmeyerCool11}. These systems also possess the largest ratio of the coupling versus cavity decay $\kappa$, viz.~$g_0/\kappa \sim 0.2$ \%. This is still two and half orders of magnitude away from the strong coupling threshold. Therefore, apart from microscopic motional degrees of freedom \cite{Esslinger08Science,Stamper2010PRL}, the cavity has to be strongly irradiated up to a high photon number $n_c \gg 1$ in order to crank up the total effect. This, however, linearizes the interaction and creates two linearly coupled linear oscillators which is a somewhat limited system. The next goal is thus to reach $g_0 \gtrsim \kappa$ where some quantum phenomena become observable \cite{RablStrong,MarquardtStro2012,PainterStrong,Nunnenkamp2}.

\begin{figure*}[htp]
 \includegraphics[width=17cm]{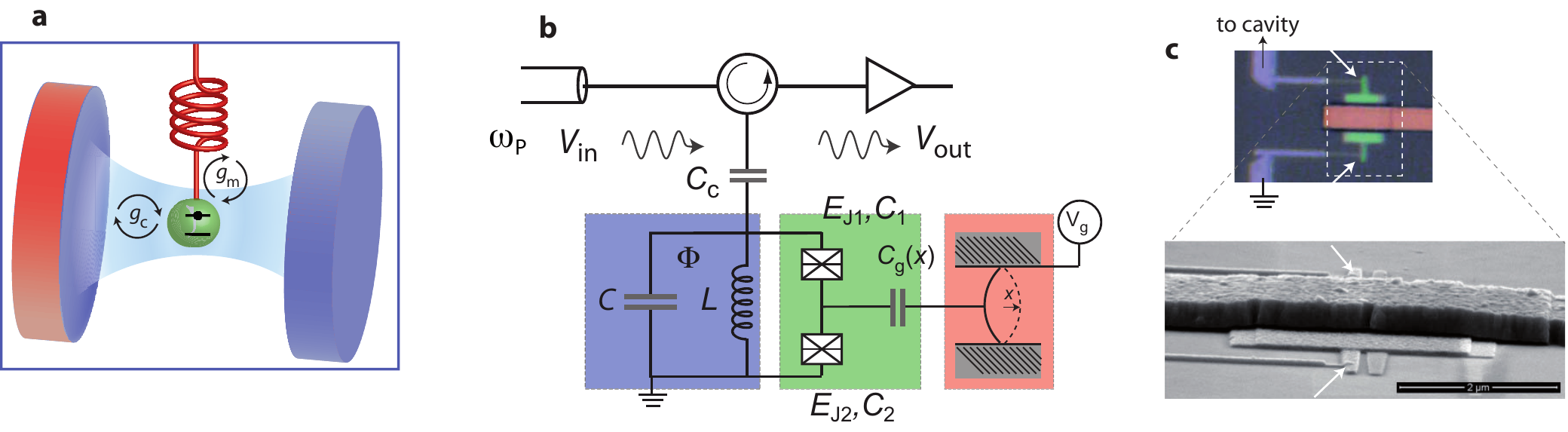}
  \caption{\emph{Optomechanics with a qubit}. (a), Illustration showing the idea for introducing a two-level system (qubit, or artificial atom) into a cavity optomechanical setup. Inside the cavity (blue) there is a quantum two-level system (green) which is mechanically compliant (red). (b), Equivalent electrical circuit of the microwave design with the same color codes as in (a). The parameters in the equivalent circuit are $C \simeq 0.33$ pF, $L \simeq 3.2$ nH, $C_g \simeq 1.8$ fF, $C_{\Sigma} \simeq 4.9$ fF, $E_J = E_{J1} + E_{J2} \simeq 10.5$ GHz, $E_C \simeq 5.3$ GHz. The reflected pump microwave $V_{\m{out}}$ is detected outside the cryostat. The Josephson junctions forming the qubit are marked by crossed boxes, and the mechanical resonator is coupled to the qubit island via a movable capacitance $C_g(x)$ (c), Micrographs of the device showing  the qubit and the mechanical resonator. The Josephson junctions are marked by arrows. }
   \label{fig:scheme}
\end{figure*}




Here we build on the scheme proposed in Ref.~\cite{LSETNEMSth} and demonstrate a novel microwave optomechanical device which allows high values of $g_0/\kappa$, and at the same time, displays novel phenomena unforeseen in cavity optomechanics. The device consists of a tripartite system made with patterned thin films of superconducting Aluminum on a silicon chip, containing first of all a microwave resonator. Another part is a micromechanical resonator, and the third is a charge qubit which consists of small Josephson tunnel junctions and couples to the other two (see Fig.~\ref{fig:scheme}). The bare cavity and the mechanical modes have the frequencies $\omega_c^0$ and $\omega_m^0$, and their dynamical variables are the creation and annihilation operators $ a^{\dag}$, $a$ and $ b^{\dag}$, $b$ respectively. The qubit is represented by the Hamiltonian $H_{\rm QB}= -B_x \sigma_x/2-B_y \sigma_y/2-B_z \sigma_z/2$. Here, $\sigma_j$ are Pauli matrices, and $B_j$ are the controllable pseudo magnetic fields which also depend on the the cavity and mechanical coordinates. The level spacing of the qubit is $B = \sqrt{\sum_{j} B_j^2}$.

A fundamental reason to expect rich optomechanical physics in the scheme owes to the nonlinearity of the quantum two-level system \cite{Etaki2008,LaHaye2009,OConnel_Nature_2010,transmonnems,Kippenberg2013TLS,CiracTLS2013,LSETNEMSth,Treutlein2014,Delsing2014}. An important ingredient here is the large dc voltage bias $V_g$ applied to the resonator represented as a movable capacitance $C_g (x)$. If coupled to a two-level system (qubit), this creates a longitudinal coupling $g_m (b^{\dag} + b) \sigma_z$  in the qubit's eigenbasis, with a maximal effect on the energy of the qubit by the mechanical resonator. When coupling this qubit to a cavity with the transverse interaction $g_c (a^{\dag} + a) \sigma_x$, the cavity frequency will experience a Stark shift $g_c^2 / (\omega_{c}^0 - B)$ which is contributed by the mechanics. The system then closely approximates the radiation-pressure interaction between the effective cavity mode (frequency $\omega_c$) and the effective mechanical mode (frequency $\omega_m$) which both are Stark-shifted from the bare frequencies. The radiation-pressure coupling strength between these modes can be amplified in this setting by the factor $C V_g/(2e)$ which amounts to about 4 to 6 orders of magnitude for typical experimental parameters  \cite{sillanpaa04}. Moreover, there are intriguing nonlinear corrections to the basic setup, first of all, a cross-Kerr type interaction in the next higher order. Most importantly, the damping of the mechanics becomes affected by the nonlinearities.






The device (see Fig.~\ref{fig:scheme} (c)) is fabricated by standard electron-beam lithography. The cavity is a 5 mm long superconducting microstrip resonating at the (bare) frequency $\omega_{c}^0 /2 \pi = 4.93$ GHz. The charge qubit has the island capacitance $C_{\Sigma} = C_{g} +C_1+C_2$ which gives the charging energy of a single electron $E_C = e^2/2C_{\Sigma}$. It is of the same order as the Josephson energy $E_J$, viz.~$E_J / E_C \simeq 2.0$. A strong qubit-cavity coupling with the coupling energy up to $\sim 500$ MHz (depending on the qubit bias) sets the system near the ultrastrong coupling \cite{Deppe2010} of circuit QED, and hence gives rise to pronounced Stark shift of the cavity. The micromechanical resonator is made as a 2 micron wide bridge suspended across 50 nm vacuum gap atop the qubit island \cite{transmonnems}.

\begin{figure}[htp]
 \includegraphics[width=8cm]{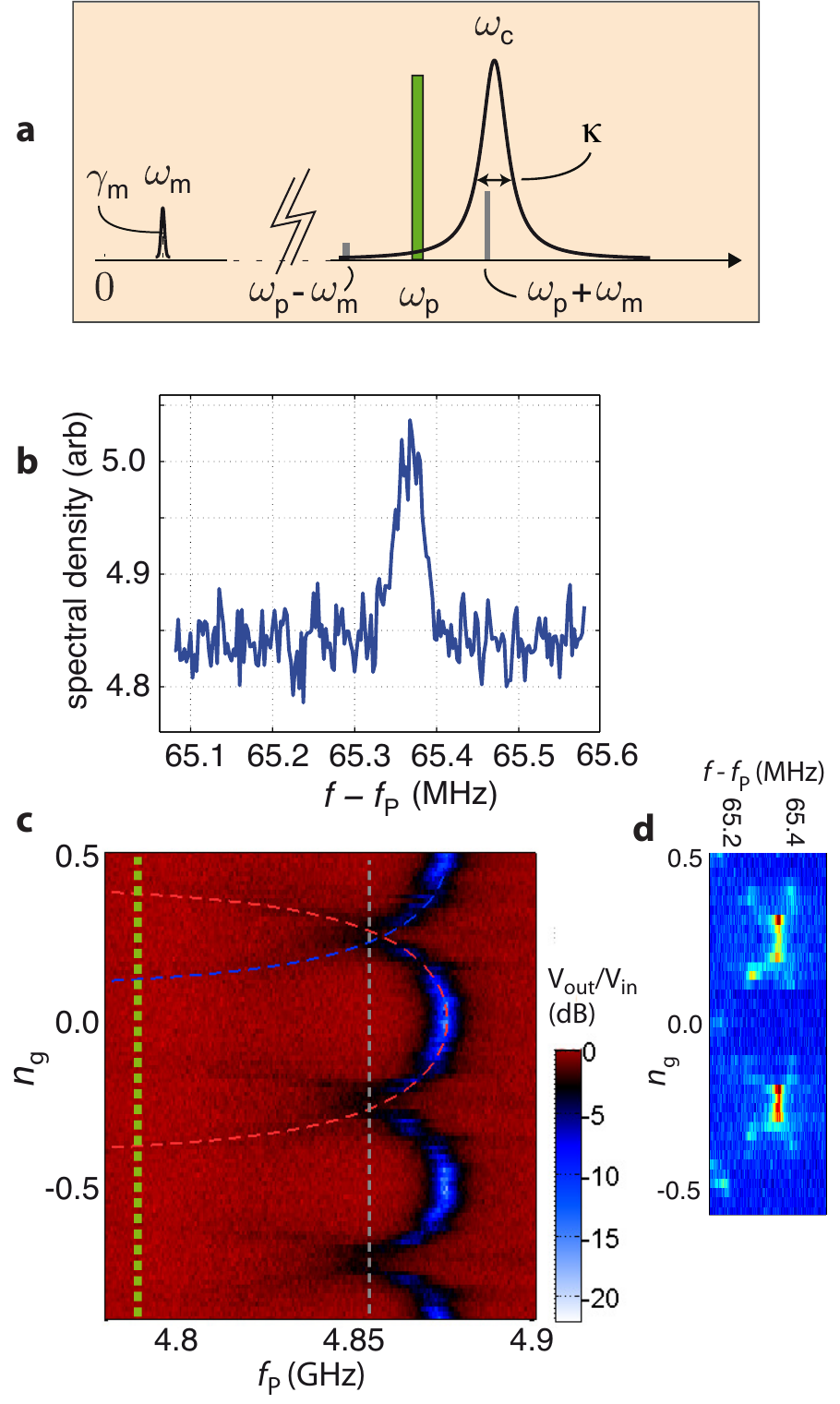}
  \caption{\emph{Basic characterization}. (a), The frequencies involved: The vertical axis gives the spectrum of either cavity ($\omega_c$) or the mechanical resonator ($\omega_m$). The pump (frequency $\omega_p$) is applied around the detuning $\Delta \simeq -1$ below the cavity frequency. (b), Thermal motion measured at 20 mK. The parameters are $\Delta \simeq -1$, $n_c \simeq 0.05$, $n_{g}  = 0.25$, $V_{g} = 6.5$ V.  (c), Gate charge modulation of the cavity resonance absorption. The resonance $\omega_c(n_g)$ of the effective cavity  appears as the periodic black-blue line. The dashed red and blue lines are theoretical fits to two of the branches, obtained by using the qubit-cavity Hamiltonian \cite{LSETNEMSth} in numerical diagonalization. The flux bias was $\Phi/\Phi_0 \simeq 0.39$. The vertical green and gray lines represent the frequencies $\omega_p$ and the  sideband $\omega_p + \omega_m$, respectively, used to obtain the data in Fig.~\ref{fig:gammaeff} (a) and (b). (d), Spectral density of the emission from the cavity around the mechanical sideband.}
\label{fig:cavity}
\end{figure}

In all measurements on the mechanical motion we use the basic idea of cavity optomechanics shown in Fig.~\ref{fig:cavity} (a), where the cavity is irradiated by the pump microwave at a frequency $\omega_p$ somewhat near the red sideband $\omega_c - \omega_m$. The pump detuning $\Delta \equiv (\omega_p - \omega_c)/\omega_m$ is hence around $\Delta \sim -1$. The motional sidebands at $\pm \omega_m$ about the pump give information about the mechanics. In the following, the sideband closer to the cavity is studied.  The main result of the radiation-pressure interaction is the ''optical spring'' which modifies both the mechanical frequency and damping. The changes exerted are denoted as $\gamma_m^{\m{opt}} \sim 4 g_0^2 n_c /\kappa$ and $\delta \omega_m^{\m{opt}}$, respectively.

 The experiments are performed in a dilution refrigerator at 20 mK temperature. We first identify the mechanical resonance by pumping the cavity very strongly up to $n_c  \sim 10^9$ such that the Josephson effect essentially averages out and leaves behind a linear cavity of the bare frequency $\omega_{c}^0$. The estimated bare optomechanical coupling between the bare cavity and the mechanics is only $g_0/2 \pi \sim 1$ Hz. This way we determine the intrinsic mechanical frequency $\omega_m/2\pi \simeq 65$ MHz (depending on $V_g$) and linewidth which is $\gamma_m/2\pi = 15$ kHz \cite{supplement}.


Next, we radically decrease the pump power by about 10 orders of magnitude down to $P_p \sim 1$ fW, however, doing so  enhances the detection sensitivity and allows us to observe the sideband peak corresponding to the thermal motion, shown in Fig.~\ref{fig:cavity} (b), at 20 mK temperature, amounting to about 6 phonons in equilibrium. The peak frequency is somewhat shifted from the calibration experiment, and the linewidth of this peak is about twice the intrinsic linewidth. As discussed below, this results from the qubit-induced nonlinearities. We now turn into careful characterization of the novel type of optomechanical interaction.


In order to measure the thermal motion, a large dc voltage $V_{g} \sim 5 ... 10$ V which establishes the interaction is applied to the mechanical resonator. On top of that, we use a fine tuning voltage in order to adjust the physics by tuning the gate charge $n_g = C_g V_{g} /2e$ (the large $V_g$ gives an irrelevant offset). 
The value of $n_g$ also affects the frequency of the effective cavity as seen in the characterization shown in Fig.~\ref{fig:cavity} (c) because of the presence of the charge-sensitive qubit component in the cavity.
The period of the cavity response is one electron ($0.5$) instead of two as expected. This is due to a quasiparticle slowly jumping on and off the island \cite{SchoelkopfTransmonB,transmonnems}. We thus see a double image, shifted by one electron, of the cavity transition. 
At a given $n_g$, there are hence two different cavity frequencies. Out of these, only one is active at a given moment, and the response is the time average of the two.

From the geometry we find $\partial C_g/\partial x \simeq 24$ nF/m and $x_{\m{zp}} \simeq 6$ fm. We determine the coupling using $g_0 =  (\partial \omega_c /  \partial x )x_{\m{zp}} $, where $(\partial \omega_c /  \partial x ) = V_g/2e (\partial \omega_c /  \partial n_g ) (\partial C_g /  \partial x ) $ can be related to the slope of the gate charge modulation of the cavity frequency seen in Fig.~\ref{fig:cavity} (c). For $n_{g} = 0.25$ where the cavity is reasonably well visible, we have $(\partial \omega_c /  \partial n_g ) \simeq (2 \pi) \cdot 250$ MHz, and hence $g_0/2\pi \simeq 0.5$ MHz when $V_g = 4.6$ V. Since $|g_0(n_g=0.25) |= |g_0 (n_g=0.75)|$, this value is the same for both cavity branches.

One can characterize the radiation-pressure physics as a function of, for example, the value of the gate charge $n_g$. Changes in $n_g$ change the value of $g_0$ which is proportional to the slope $(\partial \omega_c /  \partial n_g )$. However, a change in $n_g$ has other consequences, too. With a given pump frequency, the pump detuning $\Delta$ changes according to the effective cavity frequency. Similarly, also the effective mechanical frequency $\omega_m$ will change (due to Stark shift) with $n_g$. Finally, these effective modes exhibit the radiation-pressure interaction visible as further changes in frequency and damping due to the ''optical spring''. We denote the resulting total frequency of the mechanics as $\omega_m^{\m{tot}} =  \omega_m + \delta \omega_m^{\m{opt}}$. The total damping is given by $\gamma_m^{\m{tot}} = \gamma_m + \gamma_m^{\m{opt}} + \gamma_m^{\m{\phi}}$, where the last term $\gamma_m^{\m{\phi}}$ is due to the qubit and is discussed below. In Fig.~\ref{fig:cavity} (d) we show a result of this kind of a measurement, allowing us to connect the thermal emission peak with the cavity response. A maximum emission occurs at $n_g \simeq 0.25$ which here coincides with the sideband resonance and hence the expected maximum signal. The two crossing branches are associated with the two cavity branches.

\begin{figure*}[htp]
 \includegraphics[width=17cm]{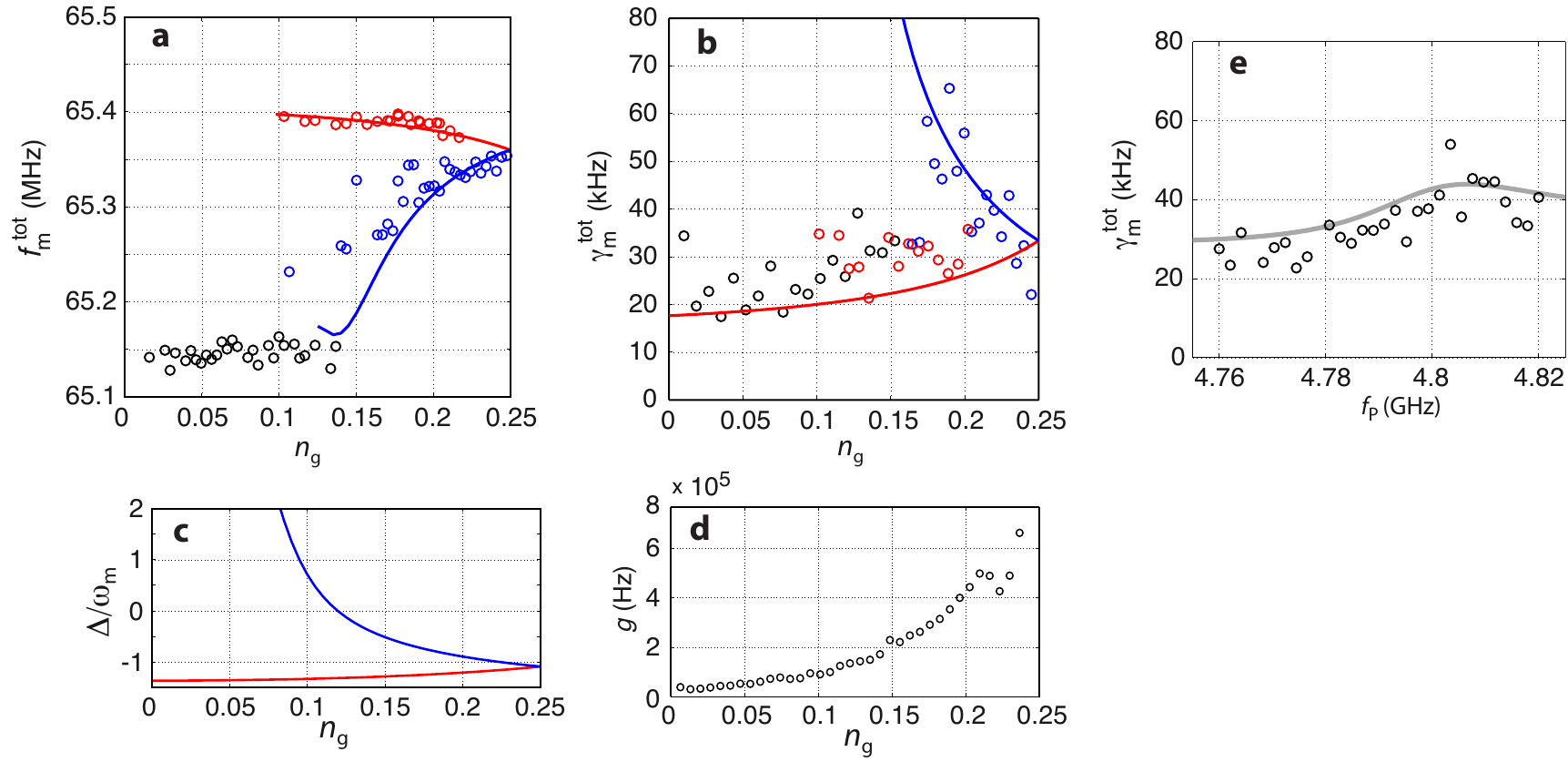}
  \caption{\emph{Optomechanics enhanced by a qubit}. (a) and (b) Optical spring as a function of gate charge. The solid lines are a fit to theory and are plotted in the region where detuning and coupling are such that the peaks are visible. The red and blue colors refer to the two cavity branches as in Fig.~\ref{fig:cavity}. The pump frequency $\omega_p/2\pi = 4.79$ GHz, and the sideband $\omega_p + \omega_m = 4.855$ GHz. (c) Illustration of the pump detuning for the two branches for the measurements in (a) and (b). (d) Radiation-pressure coupling extracted from $\omega_c(n_{g} )$. (e) Optical spring as a function of pump detuning, at fixed gate charge $n_{g}  = 0.25$. The solid line is a fit to theory. In all data, the cavity photon number is  in the range $0.1 ... 1$. This is estimated based on the input power and cable attenuation, and in the end fine-tuned as an adjustable parameter. In all the panels, $V_{g} = 4.6$ V.}
\label{fig:gammaeff}
\end{figure*}

We next look  in detail at the  frequency and linewidth of the thermal motion peak in the measurement as in Fig.~\ref{fig:cavity} (d). The extracted parameters are displayed in Fig.~\ref{fig:gammaeff} (a), (b) revealing three distinct sets of peaks. In order to understand this data, we first calculate the Stark-shifted $\omega_m$ and $\omega_c$, as well as $g_0$ from the diagonalization of the qubit-cavity Hamiltonian \cite{supplement}, and then apply the results in Refs.~\cite{LSETNEMSth,Marquardt2007} to obtain a prediction for the optical spring. Each of the two cavity branches causes its own sideband response. These arise via the detuning and coupling shown in Fig.~\ref{fig:gammaeff} (c) and (d). The blue curves in reality originate from $n_g$ values between $0.25 ... 0.5$, but reflect to the left about $n_g = 0.25$ due to the quasiparticle. The data and theory are marked with the compatible red and blue color codes in Figs.~\ref{fig:cavity}, ~\ref{fig:gammaeff}. We obtain a good agreement between the experiment and theory for the mentioned two branches.
However, the peaks visible around $n_{g}  = 0 ... 0.15$, marked in black, are not directly explained by this model. We attribute them to that the qubit spends some part of the time in the excited state as observed in this kind of measurements \cite{transmonnems}. The excited state gives rise to another  cavity frequency above $\omega_c$, having the highest $\partial \omega_c / \partial n_g$ around $n_g = 0$. The qubit being excited, however, is beyond the current model and we cannot make a quantitative comparison to theory.

\begin{figure}[ht]
 \includegraphics[width=6cm]{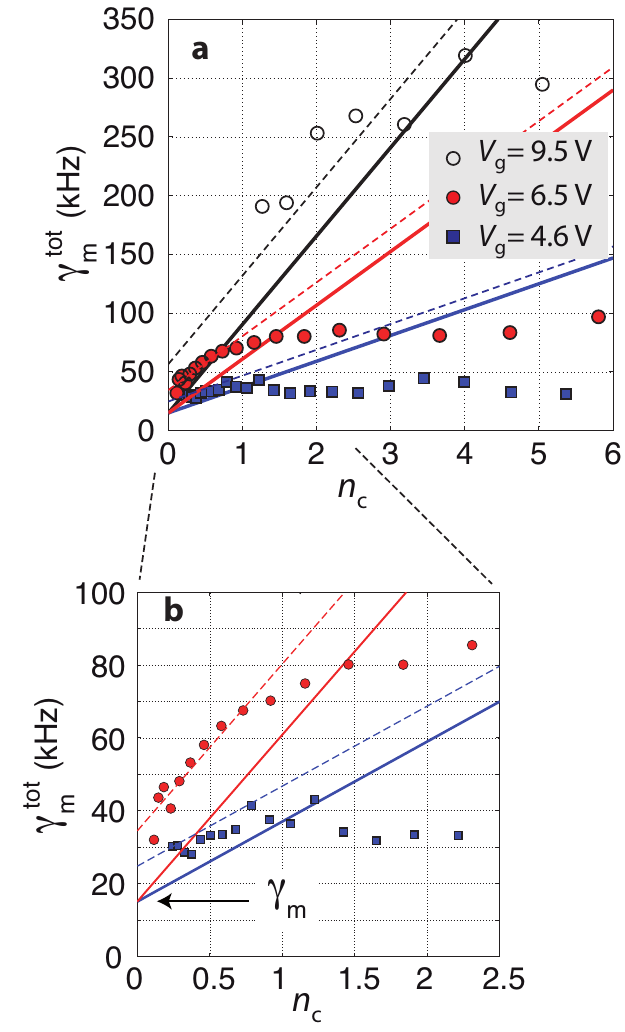}
 \caption{\emph{ Nonlinear cavity damping}. (a) The total mechanical linewidth as a function of cavity photon number. Red circles: $V_{g} = 6.5$ V, $g_0/2\pi \simeq 0.7$ MHz.  Black circles: $V_{g} = 9.5$ V yielding higher $g_0/2\pi \simeq 1.0$ MHz. Squares: smaller $g_0/2\pi \simeq 0.5$ MHz obtained with $V_{g} = 4.6$ V. The solid lines are expectations from the basic linear model, i.e.~$\gamma_m^{\m{tot}} = \gamma_m + \gamma_m^{\m{opt}}$. The dashed lines include the qubit-induced extra linewidth, $\gamma_m^{\m{tot}} = \gamma_m + \gamma_m^{\m{opt}} + \gamma_m^{\m{\phi}}$. (b), Zoom-in of the linear regime of (a). In all the data, $\Delta \simeq -1$, $\Phi/\Phi_0 \simeq 0.39$.}
\label{fig:nonlin}
\end{figure}

In the rest of the paper, we select a fixed value $n_{g}  = 0.25$ which simplifies the analysis since there is only one cavity branch and a constant $g_0$. Another way to examine the radiation-pressure interaction is to change the pump detuning (Fig.~\ref{fig:gammaeff} (e)). The total damping is maximized close to the sideband resonance in agreement with theory. 

The total damping should increase linearly with the pump photon number $n_c$ owing to the contribution $\gamma_m^{\m{opt}}$, moreover, it should depend quadratically on $g_0$ at a given $n_c$. We test this in Fig.~\ref{fig:nonlin} by varying the photon number at a few fixed values of $g_0$. The latter are set by changing the gate voltage $V_g$. As observed in Fig.~\ref{fig:nonlin}, the damping towards $n_c \rightarrow 0$ is clearly higher than the intrinsic $\gamma_m$. The extra damping is attributed to that the qubit opens another dissipation channel due to hybridization of the qubit and the mechanics \cite{LSETNEMSth}, analogously to the Stark shift, and is given by
\begin{equation}
\gamma_m^{\m{\phi}} \simeq \frac{\gamma_{ {\rm QB}}  g_{m}^2B_{1}^2}{B^3 \omega_{m} }\,.
\end{equation}
Here, $\gamma_{ {\rm QB}}$ is the qubit relaxation rate. We could not independently measure it, but a good fit of the data is obtained with a reasonable value $\gamma_{ {\rm QB}} \simeq (2\pi) \cdot 60$ MHz.



Although the damping due to the basic cavity cooling ($\gamma_m^{\m{opt}} $) gives rise to back-action sideband cooling of the mechanical resonator, further theoretical work is needed to understand the effect of the qubit contribution $\gamma_m^{\m{\phi}} $ as well as the nonlinearities such as the cross-Kerr effect on the final temperature.

In Fig.~\ref{fig:nonlin} we see that the damping grows first linearly up to about $n_c \sim 1$ and then saturates. The saturation of the radiation-pressure physics around $n_c \sim 1$ is due to reaching the limits of the linear regime of the effective cavity. This is set by that total phase excursions $\sqrt{n_c + 1/2} \phi_{\m{zp}}$ should be small as compared to $2\pi$ in order not to hit the nonlinearity of the Josephson cosine function. This is analogous to the transition of a trapped ion out from the Lamb-Dicke regime \cite{supplement}. Here, the phase zero-point fluctuation is $ \phi_{\m{zp}} \sim 0.2$ which agrees with the scale what we observe.

The largest $g_0/2\pi \simeq 1.6$ MHz obtained in this work is achieved by half a  flux quantum flux bias, and setting a high gate voltage $V_g = 9.5$ V. This value is the largest ever reported in a cavity optomechanical setting. This was achieved for integrating, for the first time, a quantum two-level system into such a setup. The ratio of coupling to the cavity linewidth, about 4 \%, should be possible to further increase by designing the coherence properties of the qubit. The novel setup introduces long-sought nonlinearities into the framework of cavity optomechanics and enables their utilization, for example, in Fock-state measurements of the phonon states, or for tests of the foundations of quantum mechanics \cite{Oosterkamp2012,AspelmeyerPlanck,Paternostro2014Macr}.


\textbf{Acknowledgements} This work was supported by the Academy of Finland and by the European Research Council (240387-NEMSQED, 240362-Heattronics, 615755-CAVITYQPD). The work benefited from the facilities at the Micronova Nanofabrication Center and at the Low Temperature Laboratory infrastructure.

\newpage

\renewcommand{\thefigure}{S\arabic{figure}}
\renewcommand{\theequation}{S\arabic{equation}}

\setcounter{figure}{0}
\setcounter{equation}{0}
\begin{widetext}
\Large{Single-photon cavity optomechanics mediated by a quantum two-level system: Supplementary information}

\normalsize

\section{The system}
\label{hamilton}

\subsection{The qubit}

Let us first present the charge qubit, marked in green background in Fig.~1b in the main text. The qubit has the junction capacitances $C_1, C_2$, and the gate capacitance
\begin{equation}
C_g(x) = C_{g0} + \delta C_g(x) \,,
\end{equation}
which has a little moving part, hence $\delta C_g(x) \ll C_{g0}$. The gate charge is defined as
\begin{equation}
n_{g}(x) = \frac{V_g C_{g}(x)}{2e} \,.
\end{equation}
It is the important control parameter setting the energy difference of having zero or one Cooper pairs on the island. Without the explicit $x$-argument, we mean the offset value
\begin{equation}
n_{g} = \frac{V_g C_{g0}}{2e} \,.
\label{eq:ngbare}
\end{equation}
The charging energy of the qubit is
\begin{equation}
\begin{split}
E_C & = \frac{e^2}{2 (C_{g0}+C_1+C_2)} \,.
\end{split}
\end{equation}
The qubit has the Hamiltonian 
\begin{equation}
H_{\rm QB}=\sum_{j=1}^3 B_j \sigma_j/2,
\label{eq:chargestateHamiltonian}
\end{equation}
where $B_1=-(E_{J1}+E_{J2}) \cos(\phi/2)$, $B_2=(E_{J1}-E_{J2})\sin(\phi/2)$ and
$B_3=4 E_C(1-2 \delta n_{g0})$ are the effective magnetic fields,
and $\sigma_j$ are Pauli matrices acting on the space spanned by the Cooper-pair charge states
$|{\rm int}(n_g)\rangle$ and $|{\rm int} (n_g) + 1\rangle$, and $\phi$ is the phase difference of the superconducting order parameters across the junction. In the absence of the cavity degree of freedom, the phase would relate to an external magnetic field applied through the superconducting loop as $\phi = 2\pi \Phi/\Phi_0$. The ground state energy is $E_{\rm QB}=-\sqrt{\sum_j  B_j^2}/2 \equiv -B/2$. 

\subsection{Full system}

Next we will discuss the full tripartite system which is represented by the lumped electromagnetic element model, as shown in \fref{fig:LSETNEMS2}. A magnetic flux $\Phi$ is externally applied through the superconducting loop. This loop also forms the cavity inductance.

The $x$-dependence of gate charge gives the coupling energy of the qubit and the mechanical resonator as
\begin{equation}
g_m = \frac{4 E_C V_g}{e} \LL( \frac{\partial C_g }{\partial x} \RR) x_{\m{zp}} \,.
\label{eq:gm}
\end{equation}

For deriving the Hamiltonian, we use for the cavity node the phase $\phi$ and the Cooper-pair number $n$, respectively as the canonical coordinate and momenta. Similarly, $\phi_I$ and $n_I$ are those for the qubit island. 

We define the charging energy of the cavity:
\begin{equation}
\begin{split}
 E_{Cc} &= \frac{e^2}{2 C}  \,, \\
\end{split}
\end{equation}
which in a typical case satisfies $ E_{Cc} \ll  E_{C}$. 

\begin{figure}[htp]
 \includegraphics[width=8cm]{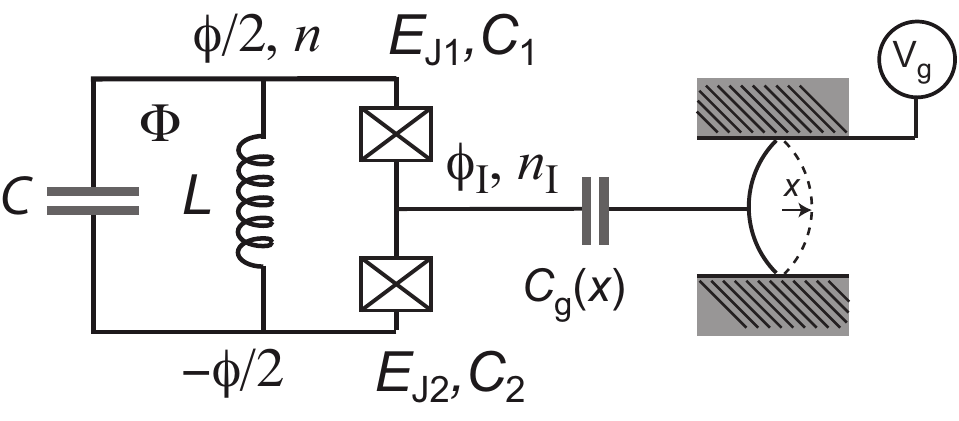}
  \caption{\emph{Circuit scheme of the cavity-qubit-mechanics tripartite system}.}
  \label{fig:LSETNEMS2}
\end{figure}

An initial form for the Hamiltonian is obtained using standard techniques as
\begin{equation}
\begin{split}
H &= 4 E_{Cc}
n^2+4 E_C \LL(n_I-n_{g}(x) \RR)^2 +\frac{\hbar^2}{4e^2}\frac{(\phi-2e\Phi/\hbar)^2}{2L} + \\ & -E_{J1}
\cos(\phi/2-\phi_I)-E_{J2} \cos(\phi/2+\phi_I) + \frac12 m\omega_m^0 x^2 + \frac{p^2}{2m}
\end{split}
\label{eq:sham1}
\end{equation}
The last two terms contain the potential and kinetic energies of the free mechanical resonator. Equation (\ref{eq:sham1}) is then quantized with the operators for the cavity ($a^{\dag}, a$) and mechanics ($b^{\dag}, b$), obtaining
\begin{equation}
\begin{split}
H &= \hbar\omega_c^0 a^{\dag} a + \hbar\omega_m^0 b^{\dag} b+4 E_C(n_I-n_{g}(x))^2 + \\&-(E_{J1}+E_{J2})\cos\big[\eta(a^{\dag}+a)+\pi \Phi/\Phi_0 \big]\cos \phi_I - (E_{J1}-E_{J2})\sin\big[\eta(a^{\dag}+a)+\pi \Phi/\Phi_0 \big]\sin \phi_I \,.
\label{eq:sham2}
\end{split}
\end{equation}
Here, $\eta=\sqrt{e^2 \sqrt{L/C}/(2\hbar)}$ is the Lamb-Dicke parameter for the cavity-qubit coupling. It has the same role as the corresponding parameter in trapped ion studies. If $\eta \ll 1$, the cavity-qubit interaction can be linearized as for a trapped ion in the Lamb-Dicke regime. In order to illustrate a simple result at this point, we further write the qubit operators in two charge-state restriction using the Pauli matrices $\sigma_i$: $\sin \phi_I = - \sigma_y/2$, $\cos \phi_I = \sigma_x/2$. We also suppose $E_{J1} = E_{J2} = E_J/2$. Equation (\ref{eq:sham2}) then simplifies to
\begin{equation}
\begin{split}
H &= \hbar\omega_c^0 a^{\dag} a + \hbar\omega_m^0 b^{\dag} b+2 E_C(1 - 2 n_g) \sigma_z - \frac{E_J}{2} \cos (\pi \Phi/\Phi_0) \sigma_x - g_m (b^{\dag}+b) \sigma_z  + \frac{E_J}{2} \eta(a^{\dag}+a) \sin (\pi \Phi/\Phi_0) \sigma_x \,.
\label{eq:sham3}
\end{split}
\end{equation}
%

The last term in \eref{eq:sham3} is the linear qubit-cavity interaction, of a typical form with the coupling energy $g_{c} \sim \eta E_J/2$. One can now get an intuitive picture of the effective cavity optomechanical system. In the full charge qubit limit, $E_J \ll E_C$, the mechanics coupling is longitudinal and shifts the qubit energy by the amount
\begin{equation}
\begin{split}
B \Longrightarrow B(x)=  B- g_m (b^{\dag}+b)  \,.
\end{split}
\end{equation}
Notice that in contrast to a regular ac Stark shift where the perturbation is perpendicular in the qubit frame, this effect is linear in $g_m$. The cavity then exhibits its own ac Stark shift due to the presence of the qubit. This shift is of the usual quadratic form since the qubit-cavity coupling is perpendicular. We define the qubit-cavity detuning $\Delta = B - \omega_c$. The shifted cavity frequency now is
\begin{equation}
\begin{split}
\omega_c^0 \Longrightarrow \omega_c^0 + \frac{g_c^2}{\Delta(x) }
= \omega_c^0 + \frac{g_c^2}{\Delta- g_m (b^{\dag}+b)}
= \omega_c^0 + \frac{g_c^2}{\Delta}+ \frac{g_m g_c^2}{\Delta^2}  (b^{\dag}+b)\,.
\end{split}
\end{equation}
The last term is the radiation-pressure coupling. In reality, the situation is more complex because of finite $E_J$, asymmetric junctions, and cavity nonlinearities neglected in \eref{eq:sham3}.


Depending on the value of the Lamb-Dicke parameter $\eta$  which can become of the order of unity, the qubit-cavity coupling can be as high as the bare cavity frequency. In this case, the system would be beyond even the ultra-strong coupling regime of (circuit) QED \cite{Deppe2010}, in the \emph{super strong} coupling regime. In the present work, we have $g_{c}/\omega_c \sim 10$ \%, so we are somewhat in the ultra-strong regime. Regarding the Lamb-Dicke regime, we have $\eta \simeq 0.1$, and hence the Lamb-Dicke limit holds when the cavity is in the vacuum state. However, the pump microwave will excite $a, a^{\dag}$ such that the driven system soon escapes out from the Lamb-Dicke regime when $a^{\dag} a \sim 1$.


The theoretical results for the energy levels used in \fref{cavityfluxdep}, and in the main text in Figs.~2c, ~3 a,b are obtained by numerical diagonalization of the ''full'' model in \eref{eq:sham2}. Notice that for the numerics, we do not need to be in the charge qubit limit which would read that junctions have the Josephson energies $E_{Ji} \lesssim E_C$, but instead we can have basically an arbitrary $E_J/E_C$ ratio. Similarly, the numerical approach goes beyond the linearized Lamb-Dicke limit of \eref{eq:sham3}.

\section{Chip layout}

 A photograph of the chip is shown in \fref{sfig:chipfoto}. The ends of the meandering inductance are close to each other such that the double-junction charge qubit can be placed to connect the ends of the meander which is required for the scheme. The device is fabricated on a regular silicon substrate using a similar process as in \cite{transmonnems}. The device was designed to not have a too small cavity capacitance in order for the system to reside in the Lamb-Dicke limit for qubit-cavity coupling (see discussion around \eref{eq:sham3}). Most of the capacitance is provided by a bonding-pad like structure visible between the coupling capacitance and the meandering inductance, marked $C$ in the figure.

\begin{figure}[th]
\centering
\includegraphics[width=9cm]{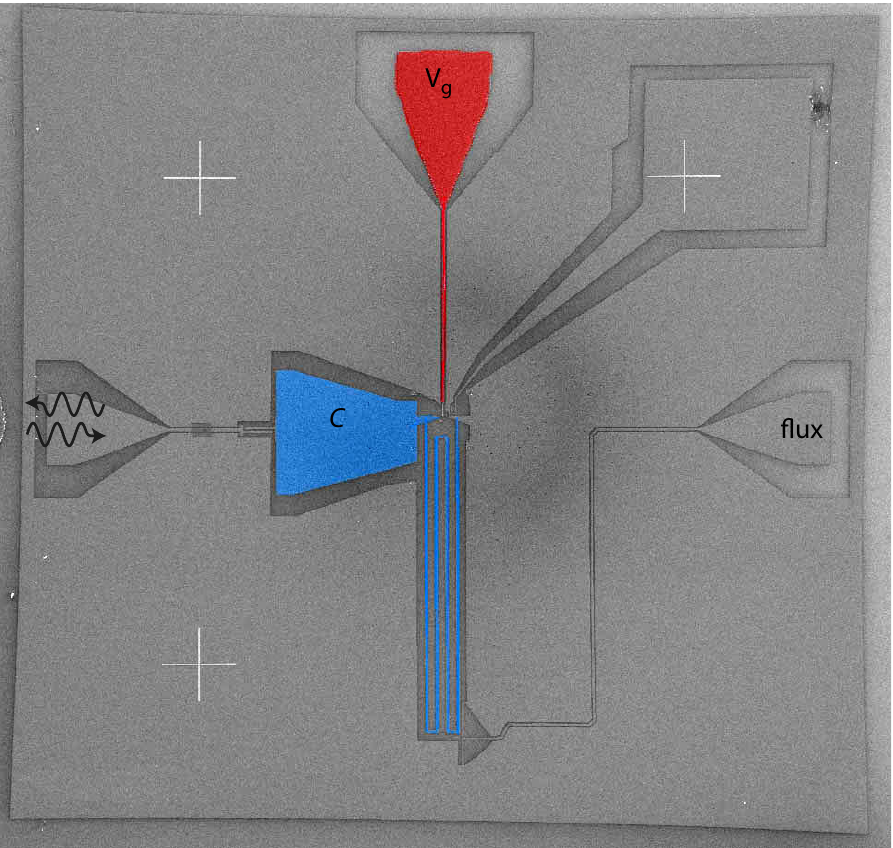}
\caption{\emph{Layout of the chip.} The gray area covering most of the chip is a ground plane. The bonding pad in the upper right corner is for a test junction situated close to the center. The cavity is marked in blue, and the connections to the mechanical resonator in red.}
\label{sfig:chipfoto}
\end{figure}

\section{Experimental setup}

The measurements are done at the base temperature of a dilution refrigerator using the cabling setup as shown in \fref{fig:samplewiring}. The pump microwave reflects from the cavity and is amplified at 4 Kelvin stage using a low-noise amplifier. The system noise temperature is in the range of 3...4 Kelvins.

Filtering the room-temperature noise away from reaching the sample via the cable connected to the mechanical resonator (i.e.,~qubit gate) is challenging because the typical solution (resistive attenuators) cannot be used owing to the high voltages involved. We hence use reflective filtering. From the qubit point of view as well, the filter needs to be reactive, because dissipative 50 $\Omega$ environment could cause too high spontaneous decay. The filtering task is carried out by a combined bias-T and low-pass filter, marked yellow in \fref{fig:samplewiring}. Without the filter, we observe substantial excitation of the qubit to the higher levels, such that the spectrum of the effective cavity changes qualitatively and cannot be modeled by ground-state transitions. With the filter installed, the excitation is modest, and cavity response well understood. The filter provides enough transmission (-33 dB) at frequencies around $\omega_m$ such that we can actuate the mechanics if needed.

\begin{figure*}[htp]
 \includegraphics[width=12cm]{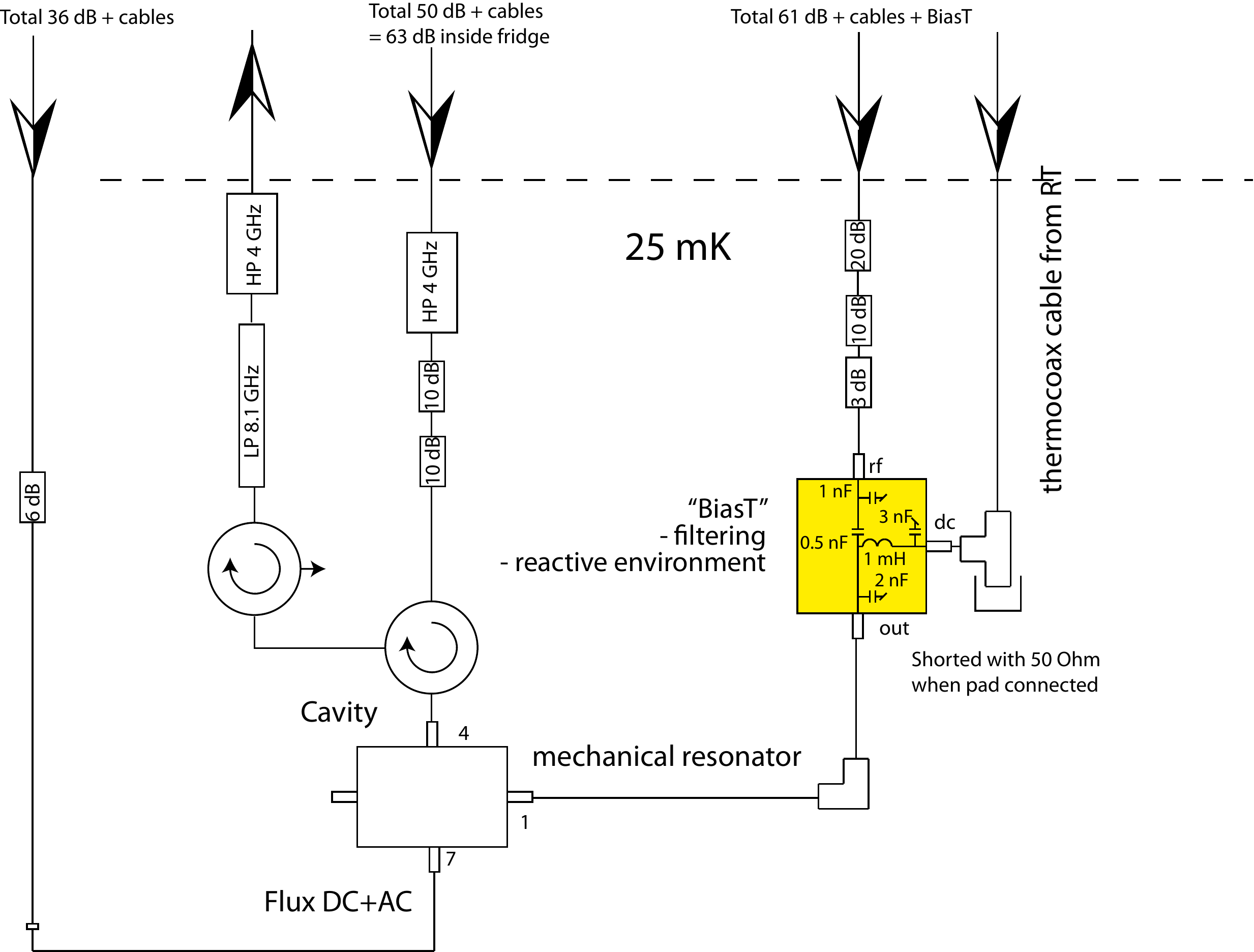}
  \caption{\emph{Experimental wiring inside the dilution refrigerator}.}
  \label{fig:samplewiring}
\end{figure*}

\section{Cavity characterization}
\label{suppcavity}

The cavity frequency has a strong dependence on the magnetic flux $\Phi$ applied through the loop, as shown in \fref{cavityfluxdep} which displays the microwave reflection response quite similar to Fig.~2c in the main text.  The cavity resonance appears as the yellow-blue colored dip varying between 4.99 GHz and 4.84 GHz, periodically in flux with the period of one flux quantum $\Phi_0$. In the measurement, we applied noise to the gate such that the response is a weighed average over all possible gate $n_g$ values. In the range $\sim 0.4 ... n_g ... 0.6$ the resonance is not well visible because of the strong gate dispersion. The theoretical curves plotted on top of the data in \fref{cavityfluxdep} are obtained by numerical diagonalization of \eref{eq:sham2}, but disregarding the mechanics (i.e. $n_g(x) \rightarrow n_g$) which here is insignificant. The overlaid curves indicate a good match between experiment and the modeling.


%
\begin{figure}[th]
\centering
\includegraphics[width=0.6\linewidth]{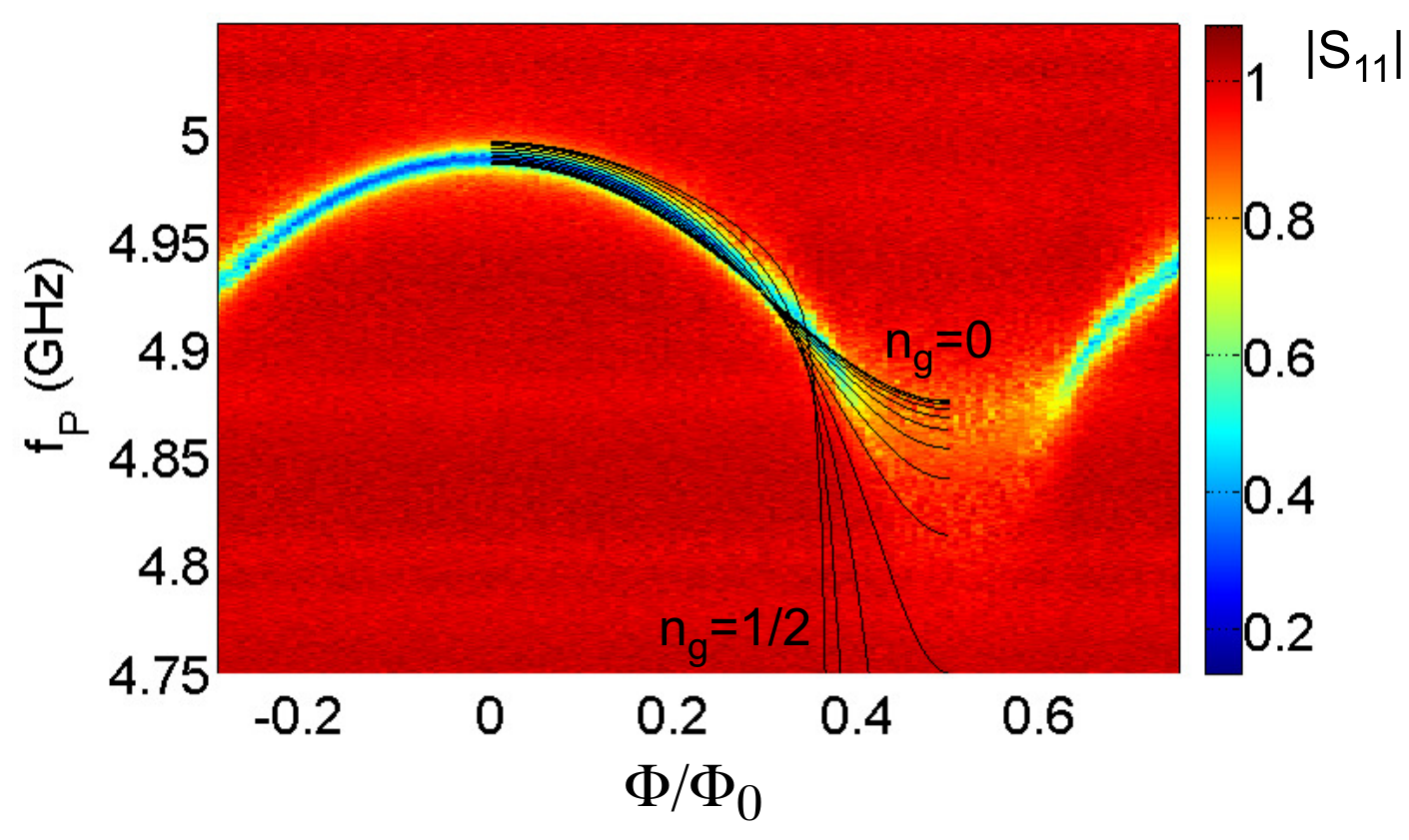}
\caption{\emph{Flux dependence of the cavity response.} The overlaid solid lines are the theoretical prediction from \eref{eq:sham2}. Each curve is plotted with a given $n_g$ value between $0 ... 1/2$ as indicated in the figure.}
\label{cavityfluxdep}
\end{figure}

The cavity linewidth $\kappa$ has a strong dependence on the qubit gate charge, as displayed in \fref{kappavsng}. While the changes are due to the qubit, we cannot make a definite conclusion of whether it is caused by qubit relaxation or dephasing. We also checked that simulating the inhomogenous broadening by establishing the theory with bare $\kappa$, and selecting a ensemble of $f_c$, and averaging over the ensemble does not dramatically change the results. 





%
\begin{figure}[th]
\centering
\includegraphics[width=0.4\linewidth]{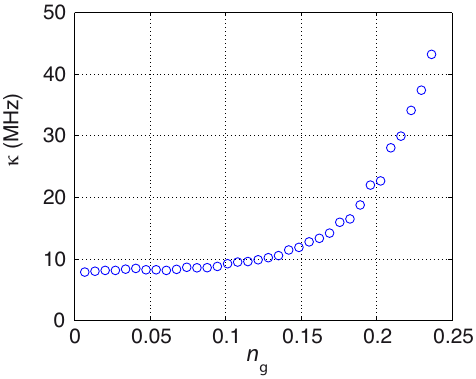}
\caption{\emph{Cavity total linewidth as a function of gate charge of the qubit.}}
\label{kappavsng}
\end{figure}

\section{Linear optomechanics at high pump power}

If the cavity photon number $n_c \gg 1$, the physics due to the Josephson junction is supposed to average out, leaving behind a linear cavity of the bare frequency $\omega_{c}^0$. This corresponds to the traditional circuit optomechanics setup.  In this regime, we can obtain an independent characterization of the mechanical resonator, with the bare radiation-pressure coupling energy $g_0/2\pi \sim 1$ Hz. Such a small coupling does not yet allow for observing the tiny thermal motion of the mechanical resonator at a low temperature. We hence actuate the resonator via the gate line, up to about 10 pm displacement amplitude, obtaining the resonance curve shown in \fref{sfig:nojos}. Notice that the mechanical frequency $\omega_m$ decreases when $V_g$ deviates from zero due to the well-known electrostatic softening of the mechanical spring.

\begin{figure}[th]
\centering
\includegraphics[width=0.4\linewidth]{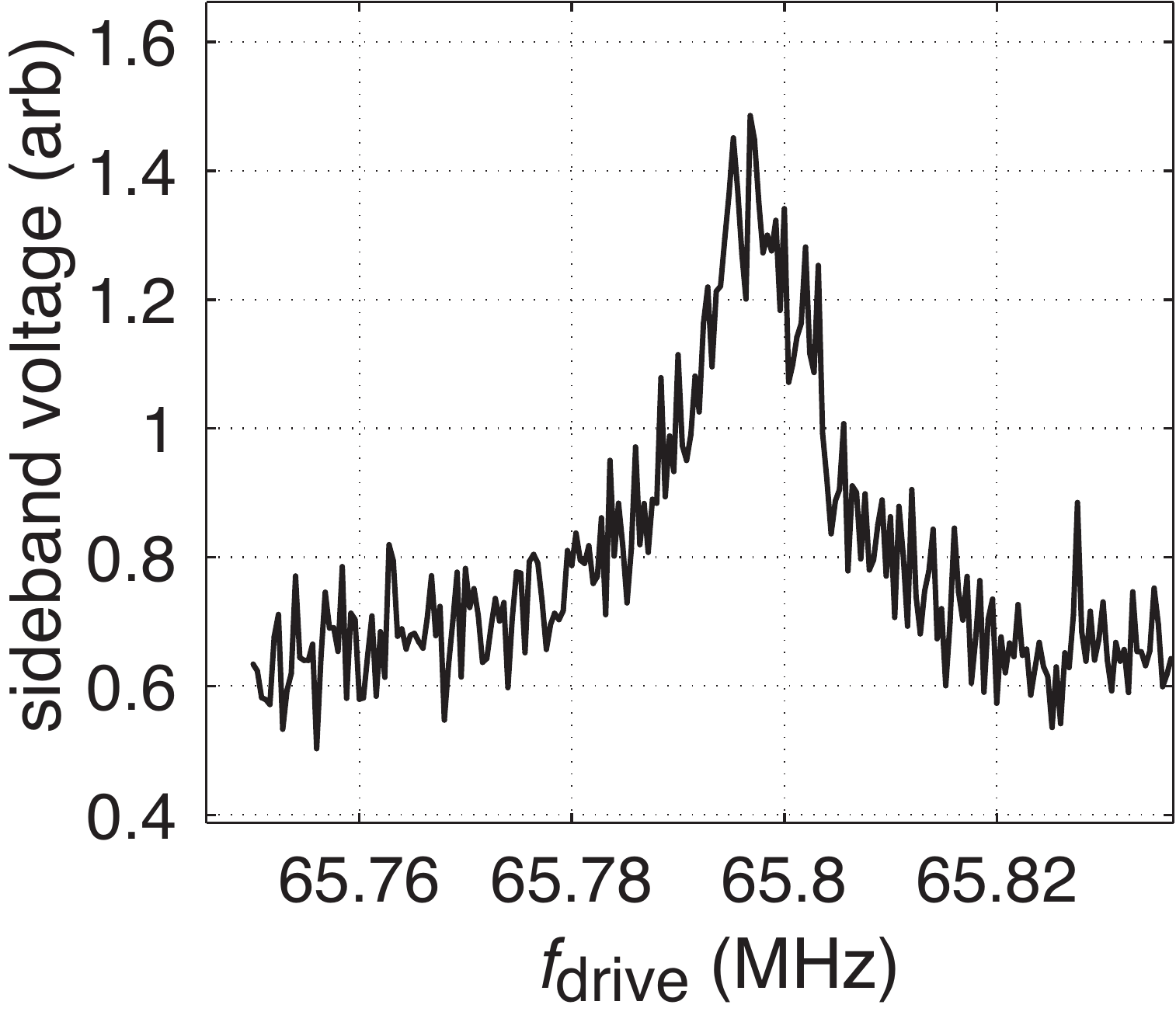}
\caption{\emph{Linear-optomechanical characterization of the mechanical resonator.} The mechanics was actuated with an ac voltage of frequency $\omega_{\m{drive}}$. The intrinsic linewidth is obtained as $\gamma_m/2\pi = 15$ kHz, and frequency $\omega_m/2\pi = 65.8$ MHz at $V_g = 1$ V.}
\label{sfig:nojos}
\end{figure}

\section{Data analysis}

In figure   \fref{fig:thermalpeaks} we display the raw data of two sets of  thermal motion sideband peaks measured under different conditions differing by the value of coupling and the cavity pump photon number. This data is used to extract the points in Fig.~4 in the main text.

\begin{figure*}[htp]
 \includegraphics[width=0.7\linewidth]{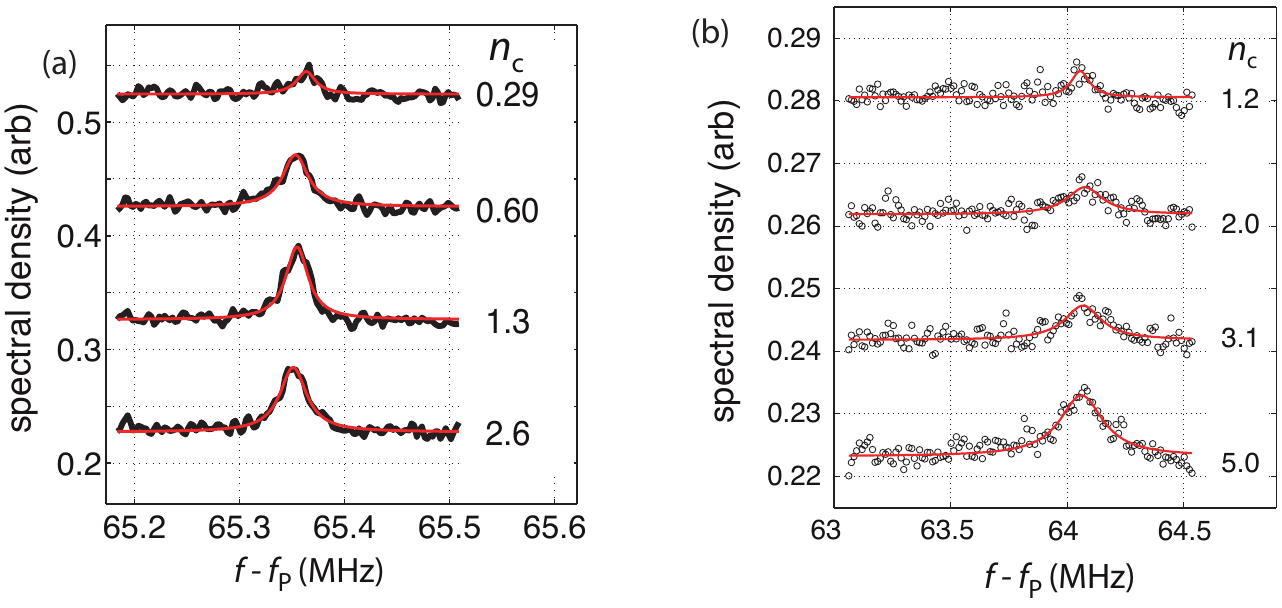}
  \caption{\emph{ Thermal sidebands}. (a), Thermal motion peaks for different cavity pump photon numbers as marked. $V_{g} = 6.5$ V, $\Phi/\Phi_0 \simeq 0.39$, $g_0/2\pi \simeq 0.7$ MHz. (b), as panel (a), but higher $V_{g} = 9.5$ V yielding higher $g_0/2\pi \simeq 1.0$ MHz. Notice the different horizontal as well as vertical scale as compared to (a).}
  \label{fig:thermalpeaks}
\end{figure*}

Let us next discuss in more detail how the effective mechanical frequencies $\omega_m^{\m{tot}}$ arise. The starting point is the ''bare'' frequency $\omega_m$. As shown in \cite{LSETNEMSth}, due to the interaction with the \emph{qubit} the mechanical resonator will exhibit an ac Stark shift
\begin{equation}
 \omega_m = \omega_m^0 - \frac{g_m^2 B_x^2}{B^3} \,.
\end{equation}
As discussed in the main text, the measured mechanical frequency is then the sum of $\omega_m$ and the contribution by the optical spring, viz.~$\omega_m^{\m{tot}} =  \omega_m + \delta \omega_m^{\m{opt}}$. One can separate the two contributions by studying the prediction for their effects. We hence plot the data in Fig.~3 but overlaid with these two contributions, in \fref{fig:mechStarkfit}. In fact, as seen in the plot, with typical parameters, $\omega_m^{\m{tot}}$ is dominated by the Stark shift. 

\begin{figure}[htp]
 \includegraphics[width=14cm]{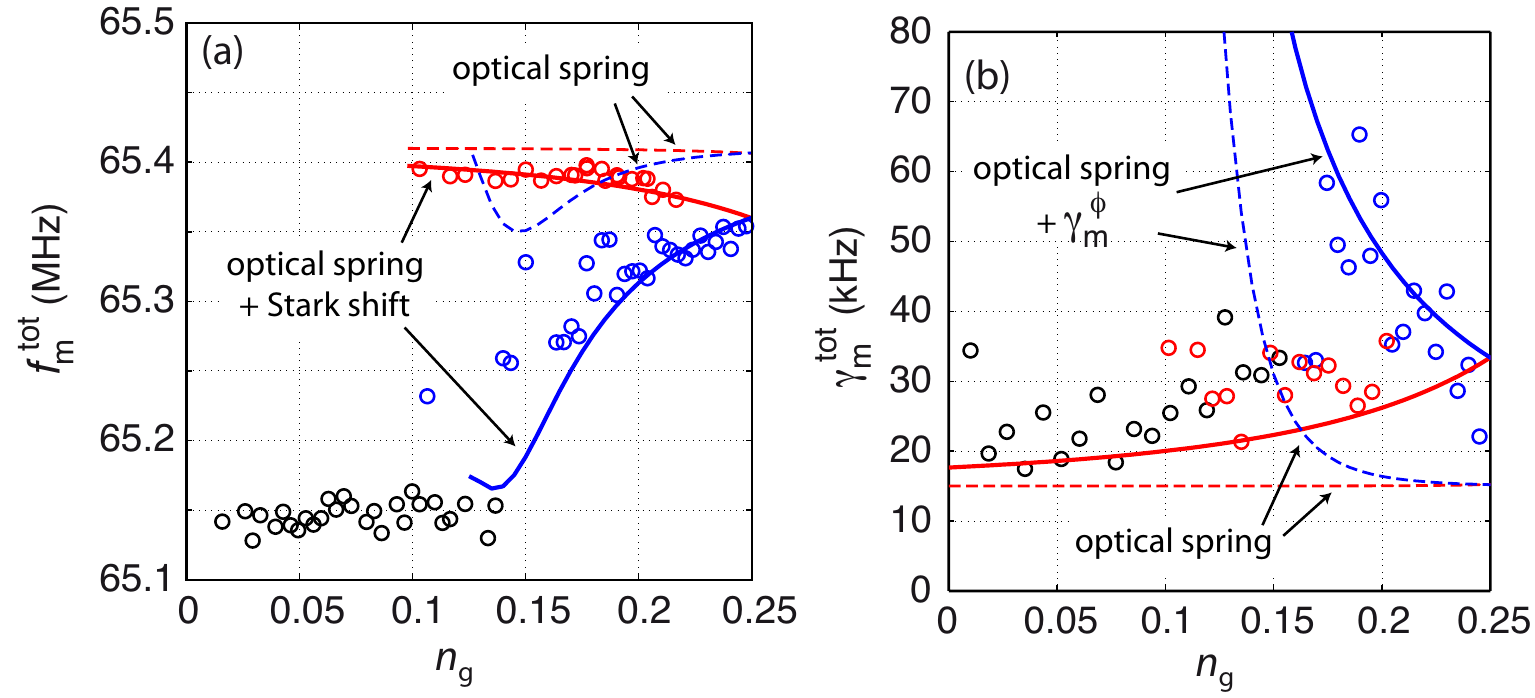}
  \caption{\emph{Modeling of the measured total mechanical frequencies}. The data is from Fig.~3a in the main text. The dashed lines indicate the contribution by the Stark shift, whereas the dotted lines are due to the optical spring.}
  \label{fig:mechStarkfit}
\end{figure}

%
%



\end{widetext}

\end{document}